\documentclass{aastex}

\RequirePackage{color}

\begin{document}

\title{Charged compact objects in the linear regime}
\slugcomment{}
\shorttitle{Short article title}
\shortauthors{Autors et al.}

\author{P. Mafa Takisa\altaffilmark{1}} \and \author{S. Ray\altaffilmark{1}} \and \author{S. D. Maharaj\altaffilmark{1}}
\affil{Astrophysics and Cosmology Research Unit,
 School of Mathematics, Statistics and Computer Science,
 University of KwaZulu-Natal, Private Bag X54001,
 Durban 4000, South Africa}


\begin{abstract}
Astrophysical compact stars provide a natural laboratory for testing theoretical models which are otherwise difficult to prove from an experimental setup. In our present work we analyse an exact solution to the Einstein-Maxwell system for a charged anisotropic compact body in the linear regime. The charged parameter may be set to zero which gives us the case of neutral solutions. We have tuned the model parameters for the uncharged case so as to match with recent updated mass-radius estimates for five different compact objects. Then we make a systematic study of the effect of charge for the different parameter set that fits the observed stars. The effect of charge is clearly illustrated in the increase of mass. We show that the physical quantities for the objects PSR J1614-2230, PSR J1903+327, Vela X-1, SMC X-1, Cen X-3 are well behaved.
\end{abstract}


\section{Introduction}\label{Sec:intro}
Exact solutions of  the Einstein-Maxwell system are of vital importance in a variety of applications in relativistic astrophysics.
 \cite{bonnor1965} demonstrated that the electric charge plays a crucial role in the equilibrium of large bodies which can possibly halt gravitational collapse. The challenge in astrophysics is to find stable equilibrium solutions for
charged fluid spheres, and to construct models of various astrophysical objects of immense gravity by considering the relevant matter distributions. Such models may successfully describe the characteristics of compact stellar objects like neutron stars, quark stars, etc.

One might argue for the occurrence of stable charged astrophysical compact objects in nature. It is true that all macroscopic bodies are charge neutral or they can have a small amount of charge  as pointed out by \cite{glendening}, so that  the structure of the star is not much affected in the latter stages of its evolution. However, there are early phases in the evolution of  compact stars, for example right at the birth from the core collapse supernova, where charge neutrality is not attained immediately and the presence of the electromagnetic field has been shown to leave a huge effect in the structure of the star. Having said so, note it has also been shown by \cite{ray2003} that from the balance of forces and the strength of their coupling, this huge charge which can disrupt the structure of the star, however leaves virtually no effect in the equation of state of the matter. Considering the protons as the carrier of the charge in the charged star, it was shown that every extra proton in a sea of $10^{18}$ baryons, can produce a total charge in the star that will change its structure.

Astrophysical compact stars are generally considered to be neutron stars. Over the past two decades, there has been considerable development in observations of compact stars. Although the mass of many of the compact stars are determined with a fair precision, the main problem comes in determining its radius. In some recent papers, improved techniques give accurate mass and radius of a few compact stars. The improved observational information about such compact stars have invoked considerable interest about the internal composition and consequent spacetime geometry of such objects.

 As an alternative to neutron star models, strange stars have been suggested in the studies of compact relativistic astrophysical bodies. Similar to neutron stars, strange stars are considered likely to form from the core collapse of a massive star during a supernova explosion, or during a primordial phase transition where quarks clump together. Another hypothesis is that an accreting neutron star in a binary system, can accrete enough mass to induce a phase transition at the centre or the core, to become a strange star. In the literature, there are numerous models of neutron stars and strange stars. In this paper we use the most widely studied strange star model, namely the MIT Bag model.
Studies of strange stars have been mostly performed within the framework of the Bag model as the physics of high densities is still not very clear. \cite{chodos1974} used the phenomenological MIT Bag model, where they assumed that the quark confinement is caused by a universal bag pressure at the boundary of any region containing quarks, namely the hadrons. The equation of state describing the strange matter in the bag model has a simple linear form from the treatment of \cite{witten1984}. \cite{weber2005} has shown that for a stable quark matter, the bag constant is restricted to a particular range.

It is remarkable to note that a strange matter equation of state seems to explain the observed compactness of many astrophysical bodies such as Her X-1, 4U 1820-30, SAX J 1808.4-3658, 4U 1728-34, PSR 0943+10, and RX J185635 as pointed out by  \cite{rahaman2012}.  \cite{dey1998} studied a new approach for strange stars by assuming an interquark vector potential originating
 from gluon exchange and a density dependent scalar potential which restores chiral symmetry at a high density. In the  \cite{dey1998}  formulation, the equation of state can also be approximated to a linear form. If pulsars are modelled as strange stars, the linear equation of state appears to be a feature in the composition of such objects as established in the analysis of \cite{sharma2007a}. For a given central density or pressure, the conservation equations can be integrated to compute the macroscopic features such as the mass and the  radius of the star.

In situations where  the densities inside the star are beyond nuclear matter density, the matter anisotropy can play a crucial role, as the conservation equations are modified. \cite{usov2004} suggested the consideration of anisotropy in modelling strange stars in the presence of strong electric field.
The analysis of static spherically symmetric anisotropic fluid spheres is important in relativistic astrophysics. Since the first study of \cite{bowers1974} there has been much research in the study of anisotropic relativistic matter in general relativity. It has been pointed out that nuclear matter may be anisotropic in high density ranges of order  $10^{15}{\rm g}~{\rm cm}^{-3}$, where nuclear interactions have to be treated relativistically, originally in the treatment of \cite{ruderman1972}. It has been noted that anisotropy can arise from different kinds of phase transitions by \cite{sokolov1980} or pion condensation by \cite{sawyer1972}. The role of charge in a relativistic quark star was considered by  \cite{mak2004}.

In the present work we study the regular exact model of the the Einstein-Maxwell system found by  \cite{takisa2013} by testing for consistency and compatibility with observations in this model.
We use this model to find the maximum mass and physical parameters of observed compact objects, namely
PSR J1614-2230, PSR J1903+327, Vela X-1, SMC X-1 and Cen X-3, which have been recently identified by  \cite{gangopadhyay2013} to be strange stars.
In Sect. \ref{Sec:model}, the Einstein-Maxwell field equations are briefly reviewed and the
 \cite{takisa2013} model is revisited. Recent observations are presented in Sect. \ref{Sec:recent}.
In Sect. \ref{Sec:uncharged}, we present and discuss our results obtained for the uncharged case
and compare them to values of masses derived from current accurate observations of compact objects.
In Sect. \ref{Sec:charged}, we apply finite charge to the uncharged systems presented in Sect. \ref{Sec:uncharged}, and observe the changes. We discuss and conclude our results in Sect. \ref{Sec:conc}.

\section{The model}\label{Sec:model}
The metric of a static spherically symmetric spacetime in curvature coordinates reads
\begin{equation}
\label{f1} 
ds^{2} = -e^{2\nu} dt^{2} + e^{2\lambda} dr^{2}
+ r^{2}(d\theta^{2} + \sin^{2}{\theta} d\phi^{2}),
\end{equation}
where $\nu=\nu(r)$ and $\lambda=\lambda(r)$.
The energy momentum tensor for an anisotropic charged imperfect
fluid sphere is of the form
\begin{eqnarray}
\label{eq:f2} T^{ab}=\mbox{diag}(-\rho -\frac{1}{2}E^2, p_r-\frac{1}{2}E^2, p_t + \frac{1}{2}E^2, p_t+ \frac{1}{2}E^2),
\end{eqnarray}  
where $\rho$, $p_{r}$, $p_{t}$ and  $E$ are the density, radial pressure, tangential pressure and electric field intensity respectively.
For a physically realistic relativistic star we expect that the
matter distribution should satisfy a barotropic equation of state
$p_r=p_r(\rho)$; the linear case is given by
\begin{equation}
\label{helen}
\label{eq:f18} p_r = \alpha\rho - \beta ,
\end{equation}
where $\beta=\alpha\rho_{\varepsilon}$. The constant $\alpha$ is constrained by the sound speed causality condition ($\alpha=\frac{dp_{r}}{dr}\leq 1$) and $\rho_{\varepsilon}$
represents the density at the surface $r=\varepsilon$.

The gravitational interactions on the matter and electromagnetic fields are governed by a relevant set of field equations.
These interactions are contained in the Einstein-Maxwell system
\begin{eqnarray}
\label{P10a}
G^{ab} &=&  kT^{ab},\\
\label{P10b}
F_{ab;c}+F_{bc;a}+F_{ca;b} &=& 0,\\
\label{P10c}
{F^{ab}}_{;b} &=& 4\pi J^{a},
\end{eqnarray}
where the coupling constant $k=8\pi$ $(G = c= 1)$ in geometrized units.
The system above is a highly nonlinear system of coupled, partial differential equations
governing the behaviour of the gravitating system in the presence of the electromagnetic field.

For static, charged anisotropic matter with the line element (\ref{f1}), the Einstein-Maxwell system  (\ref{P10a})-(\ref{P10c}) takes the form
\begin{eqnarray}
\label{f3} 
 8\pi\rho + \frac{1}{2}E^{2}&=&\frac{1}{r^{2}} \left[ r(1-e^{-2\lambda}) \right]',\\
 \label{f4} 
8\pi p_r -\frac{1}{2}E^{2}&=&- \frac{1}{r^{2}} \left( 1-e^{-2\lambda} \right) + \frac{2\nu'}{r}e^{-2\lambda} ,\\
\label{f5} 
8\pi p_t + \frac{1}{2}E^{2}&=&e^{-2\lambda}\left( \nu'' + \nu'^{2}+ \frac{\nu'}{r}\lambda' -\frac{\lambda'}{r} -\nu\right),
\label{f6}  \\
\sigma & = & \frac{1}{4\pi r^{2}} e^{-\lambda}(r^{2}E)', \label{x}
\end{eqnarray}
where $\sigma=\sigma(r)$ is called proper charge density and primes denote differentiation with respect to $r$.
We note that equations (\ref{f3})-(\ref{f6}) imply
\begin{equation}
\label{pedr}
 \frac{dp_{r}}{dr}=\frac{2}{r}(p_{t}-p_{r})-r(\rho+p_{r})\nu^{\prime}+\frac{E}{4\pi r^{2}}\left( r^{2}E\right)^{\prime},
\end{equation}
which is the Bianchi identity representing hydrostatic equilibrium of the charged anisotropic fluid. Equation (\ref{pedr}) indicates that the anisotropy
and charge influence the gradient of the pressure. These quantities may drastically affect quantities of physical importance such as surface tension
as established by \cite{sharma2007b} in the generalised Tolman-Oppenheimer equation (\ref{pedr}).
We define the gravitational mass to be
\begin{equation}
\label{mass}
m(r)=4\pi\int^{r}_{0}\left(\rho(\omega)+\frac{E^{2}}{8\pi}\right)d\omega,
\end{equation}
in the presence of charge.

In this paper, we utilise the results of \cite{takisa2013}, with a linear equation of state. The motivation for this is that their results are consistent with the
observed X-ray binary pulsar SAX J1808.4-3658. It is likely that the exact solutions of  \cite{takisa2013} may be applicable to other observed astronomical bodies.
With the equation of state (\ref{helen}), the solution to the Einstein-Maxwell system (\ref{f3})-(\ref{x}) can be written as
\begin{eqnarray}
\label{S10a}
e^{2\lambda} &=& \frac{1+ar^{2}}{1+(a-b)r^{2}},\\
\label{S10b}
  e^{2\nu} &=& A^{2}(1+ar^{2} )^{2t}[1+(a-b)r^{2}]^{2n}\nonumber\\
&&\times\exp\left[-\frac{ar^{2}[ s(1+\alpha)+2\beta ]}
{4(a-b)}\right],\\
\label{S10c}
 \rho &=& \frac{2b(3+ar^{2})-sa^{2}r^{4}}{16\pi(1+ar^{2})^{2}}, \\
\label{S10d}
p_{r} &=& \frac{\alpha[2b(3+ar^{2})-sa^{2}r^{4}]}{16\pi(1+ar^{2})^{2}}-\beta,\\
\label{S10e}
p_{t} &=& p_{r}+\Delta, \\
\label{S10f}
8\pi\Delta &=& \frac{-b}{(1+ar^{2})}-\frac{b(1+5\alpha)}{(1+\alpha)(1+ar^{2})^{2}}
 +\frac{2\beta}{1+\alpha}+\frac{r^{2}[1+(a-b)r^{2}]}{(1+ar^{2})}\nonumber\\
&&\times\{4\left(\frac{a^{2}t(t-1)}{(1+ar^{2})^{2}}\right.
 \left.+\frac{2a(a-b)tn}{(1+ar^{2})[1+(a-b)r^{2}]}\right)+\left(\frac{4(a-b)^{2}n(n-1)}{[1+(a-b)r^{2}]^{2}}\right)\nonumber\\
&&-a[s(1+\alpha)+2\beta]\times\frac{(a(t+n)[1+(a-b)r^{2}]-bn)}{(a-b)(1+ar^{2})[1+(a-b)r^{2}]}\nonumber\\
&&+\frac{a^{2}[s(1+\alpha)+2\beta]^{2}}{16(a-b)^{2}}\}-\frac{4[1+ar^{2}(2+(a-b)r^{2})]-b(5+\alpha)r^{2}}{4(a-b)(1+\alpha)(1+ar^{2})^{3}[1+(a-b)r^{2}]}\nonumber\\
&&\times[-8b^{2}n+a^{3}r^{2}(-8(t+n)+[s(1+\alpha)+2\beta]r^{2})
 +a^{2}(8(t+n)(2br^{2}-1)\nonumber\\
&&+[s(1+\alpha)+2\beta](2-br^{2})r^{2}
+a(-8b^{2}(t+n)r^{2}+[s(1+\alpha) +2\beta]\nonumber\\
&&+b(8t+16n-[s(1+\alpha)+2\beta]r^{2}))],\\
\label{S10g}
E^{2}&=& \frac{sa^{2}r^{4}}{(1+ar^{2})^{2}},\\
\label{S10h}
\sigma^{2} &=& \frac{sa^{2}r^{2}[1+(a-b)r^{2}](2+ar^{2})^{2}}{\pi(1+ar^{2})^{5}},\\
\label{S10i}
m(r)&=&\frac{1}{8}\left[\frac{ s(-15-10ar^{2}+2a^{2}r^{4})r}{3a(1+ar^{2})} +\frac{4br^{3}}{(1+ar^{2})} +\frac{5s\arctan(\sqrt{ar^{2}})}{a^{3/2}}\right],
\end{eqnarray}
where $A$, $a$, $b$ and $s$ are constants. In the above equations the constants \textit{m} and \textit{n} are given by
\begin{eqnarray}
t &=& \frac{4\alpha b-(1+\alpha)s}{8b},\nonumber\\
n &=& \frac{1}{8b(a-b)^{2}}[a^{2}((1+\alpha)s-4\alpha b) 2ab^{2}(1+5\alpha) +b^{2}(-2b(1+3\alpha)+2\beta)].\nonumber
\end{eqnarray}
The exact solution (\ref{S10a})-(\ref{S10i}) of the Einstein-Maxwell system is written in terms of 
elementary functions.

The constants $a$, $b$, $s$ have the dimension of $length^{-2}$. 
We make the following  transformations for simplicity in numerical calculations:
\[\tilde{a}=a \Re^{2},~~\tilde{b}=b\Re^{2},~~\tilde{s}=s \Re^{2},\] where $\Re$ is a parameter which has the dimension of $length$.
Based on the requirements of  \cite{delgaty1998}, we impose restrictions on our model to make it physically relevant.
The values of $\tilde{a}$, $\tilde{b}$, $\tilde{s}$ should be chosen so that:
\begin{itemize}
 \item  the energy density $\rho$ remains positive
  inside the star,
  \item the radial pressure $p_{r}$ should vanish at the boundary of the star ($p_{r}(\varepsilon)=0$),
   \item the tangential pressure $p_{t}$ should be positive within the interior of star, 
   \item the gradient of pressure $\frac{dp_{r}}{dr}<0$ in the interior of the star,
   \item At the centre $p_{r}(0)=p_{t}(0)$ and $\Delta(0)=0$,
   \item The metric functions $e^{2\lambda}$, $e^{2\nu}$ and the electric field intensity
    $E$ should be positive and non singular throughout the interior of the star.
    \item A the centre density $\rho(0)=\rho_{c}$ must be finite.
    \item across the boundary $r=\varepsilon$:
\begin{eqnarray}
 e^{2\nu(\varepsilon)}&=&1-\frac{2M}{\varepsilon}+\frac{Q^{2}}{\varepsilon^{2}},\nonumber\\
\label{S105}
e^{2\lambda(\varepsilon)}&=&\left(1-\frac{2M}{\varepsilon}+\frac{Q^{2}}{\varepsilon^{2}}\right)^{-1},\nonumber\\
m(\varepsilon)&=&M. \nonumber
\label{S106}
\end{eqnarray}
\end{itemize}

\section{Recent observations}\label{Sec:recent}
For a pulsar in a binary system,  \cite{jacoby2005} and  \cite{verbiest2008} used detection of the general relativistic Shapiro
delay to infer the masses of both the neutron star and its binary companion to high precision. Based on this approach \cite{demorest2010} presented radio timing observations of the binary
millisecond pulsar PSR J1614-2230, which showed a strong Shapiro delay signature. The implied pulsar mass of $(1.97 \pm 0.08M_\odot)$ is by far the highest yet measured with accurate precision.

 \cite{freire2011} utilised the Arecibo and Green Bank radio timing observations and included a full determination of the relativistic Shapiro delay, a very precise measurement of the apsidal motion and new constraints of the orbital orientation of the system.
Through a detailed analysis, they derived new constraints on the mass of the pulsar and its companion and determined the accurate mass for PSR J1903+0327 $(1.667 \pm 0.02M_\odot)$.

Recently  \cite{rawls2011} have found an improved method for determining the mass of neutron stars such as (Vela X-1, SMC X-1, Cen X-3) in eclipsing X-ray pulsar binaries. They used a numerical code based on Roche geometry with various optimizers to analyze the published data for these systems, which they supplemented with new spectroscopic and photometric data for 4U 1538-52. This allowed them to model the eclipse duration more accurately, and they calculated an improved value for the neutron star masses. Their derived values are $(1.77 \pm 0.08M_\odot)$ for Vela X-1, $(1.29 \pm 0.05M_\odot)$ for LMC X-4  and $(1.29 \pm 0.08M_\odot)$ for Cen X-3.

There have been similar observations for other stars, but for our present work, we restrict ourselves to these five stars only.

\begin{figure*}
\centering
\begin{tabular}{cc}
\includegraphics[width=0.52\textwidth]{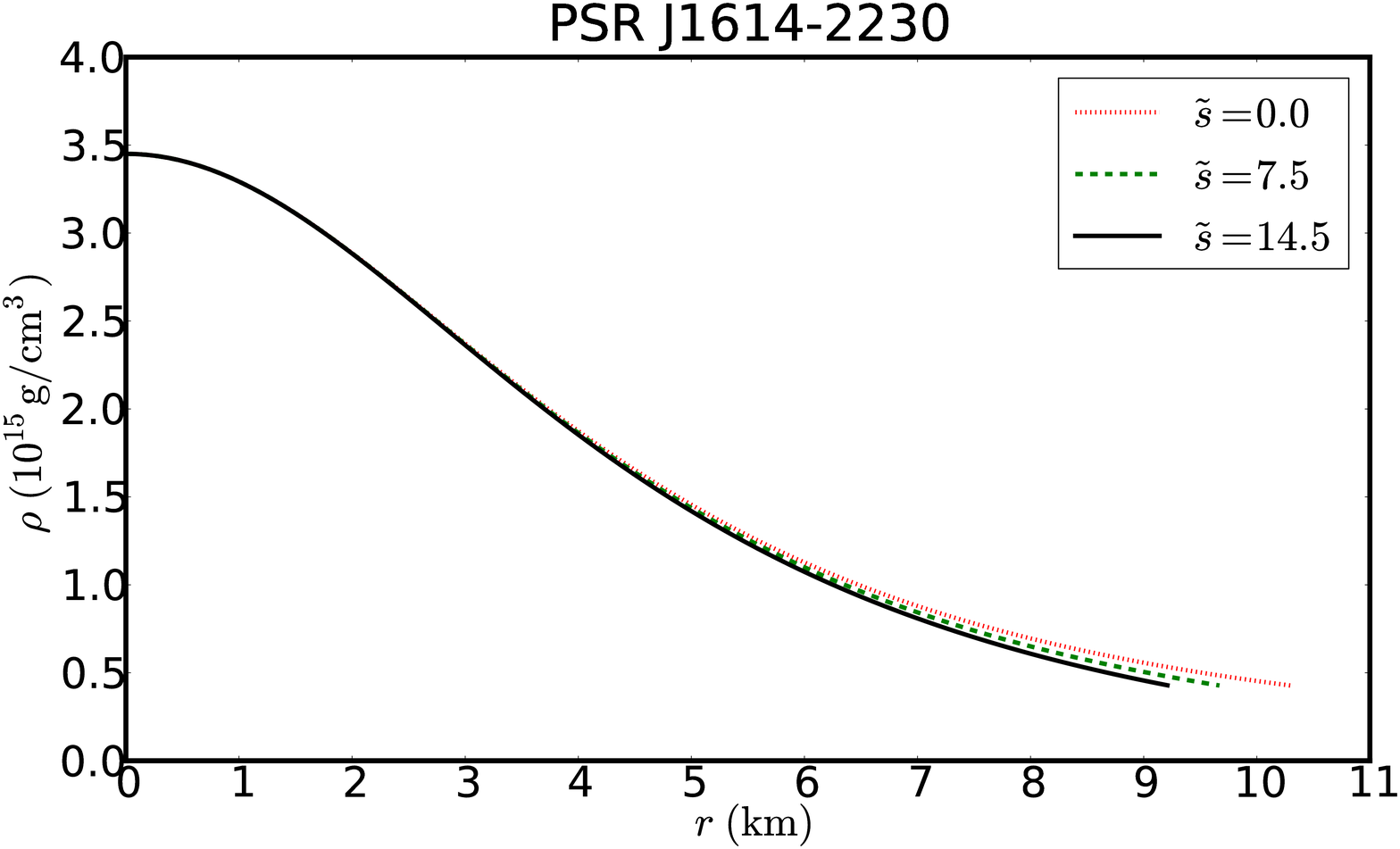} &
\includegraphics[width=0.52\textwidth]{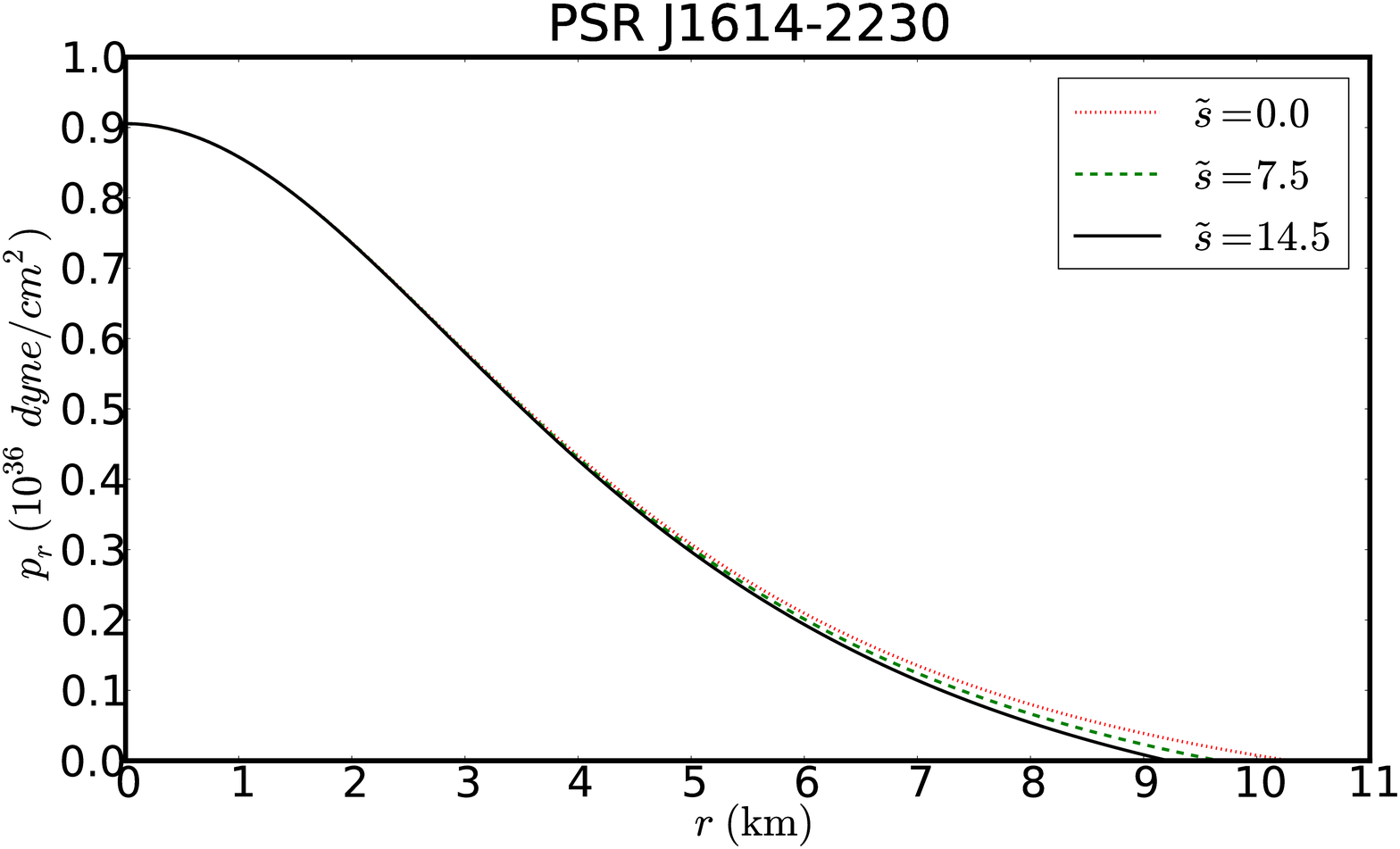}
\end{tabular}
\begin{tabular}{cc}
\includegraphics[width=0.52\textwidth]{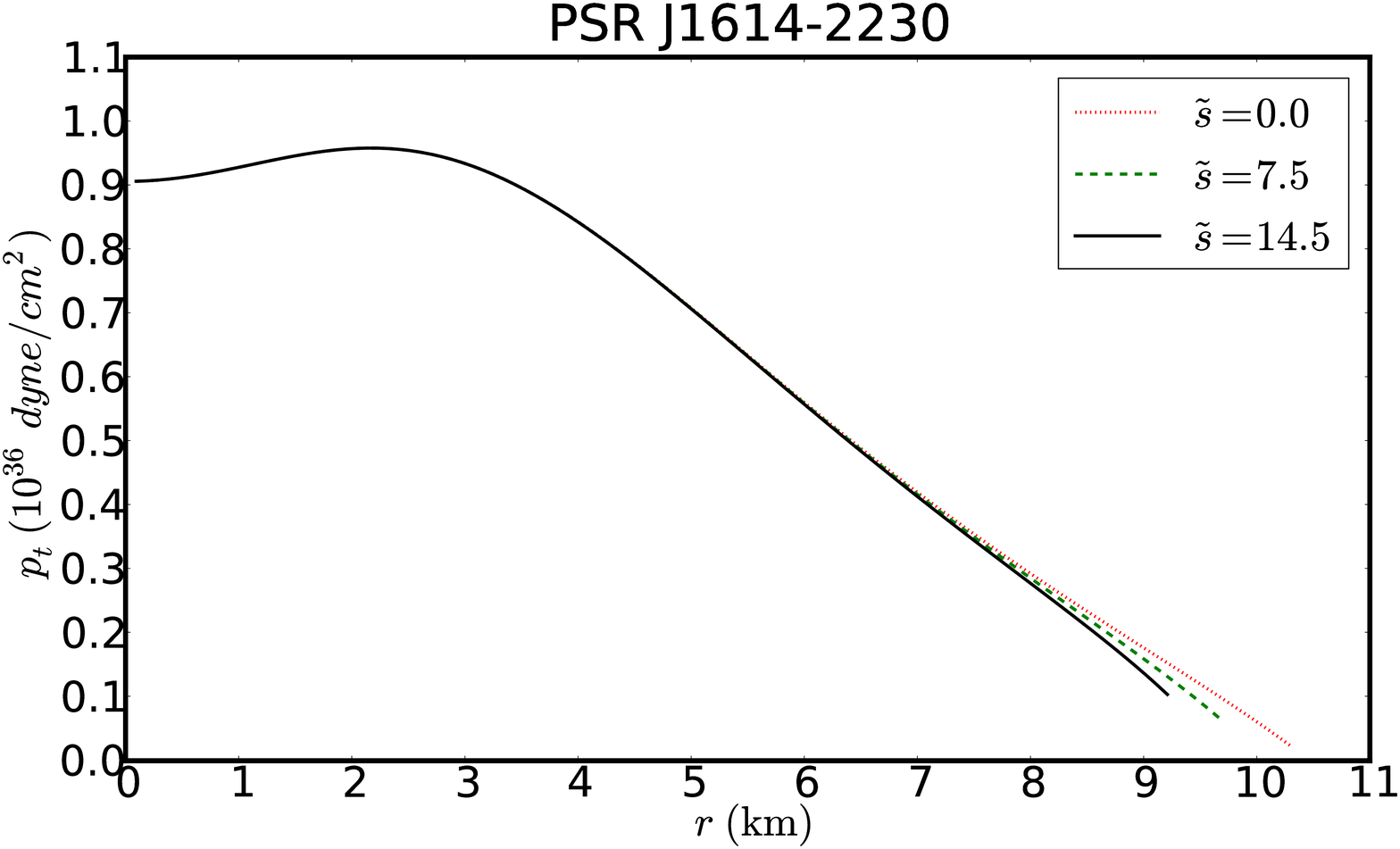} &
\includegraphics[width=0.52\textwidth]{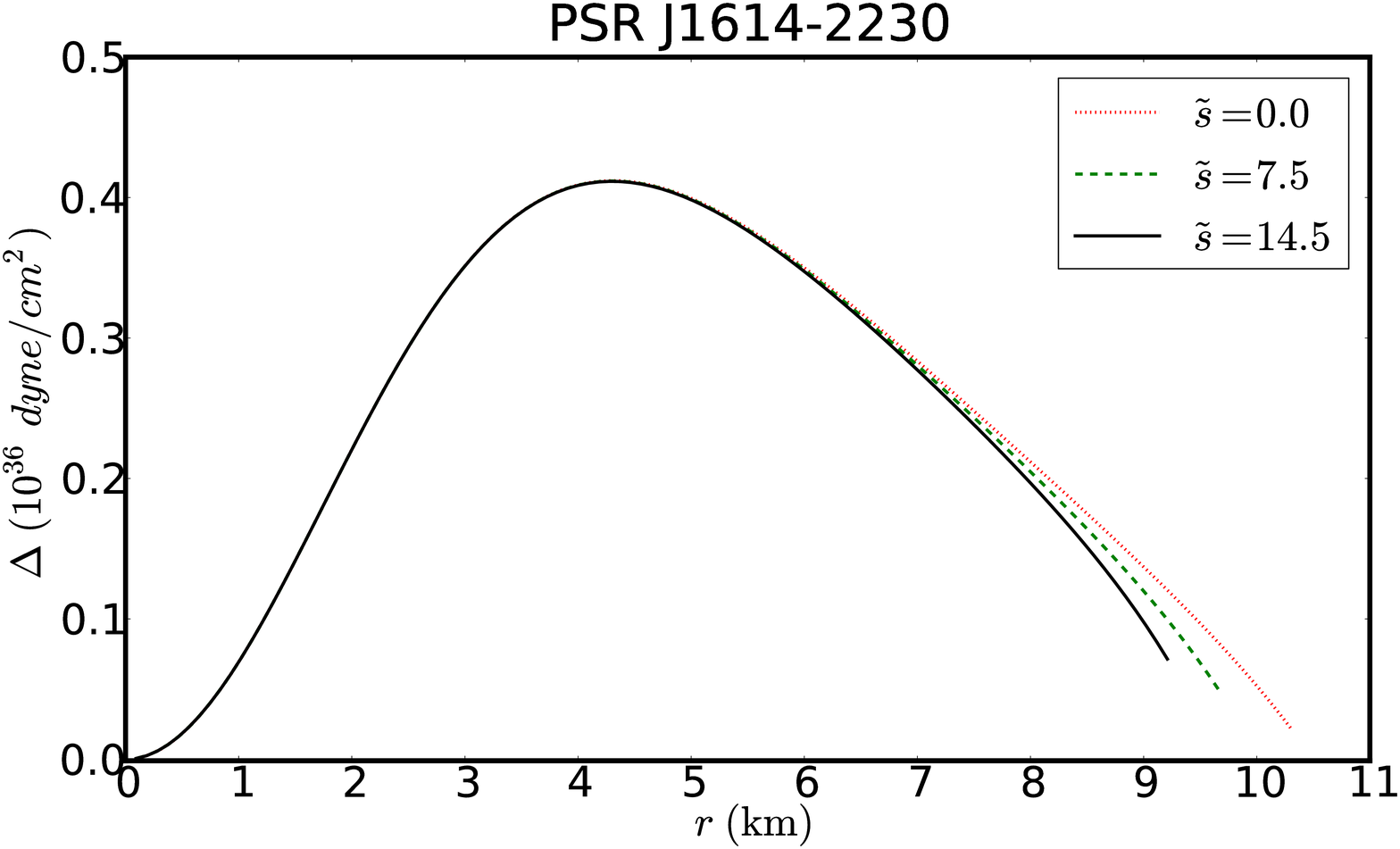}
\end{tabular}
\caption{PSRJ1614-2230, for the uncharged and charged cases.}
\label{fig:1614}
\end{figure*}

\begin{figure*}
\centering
\begin{tabular}{cc}
\includegraphics[width=0.52\textwidth]{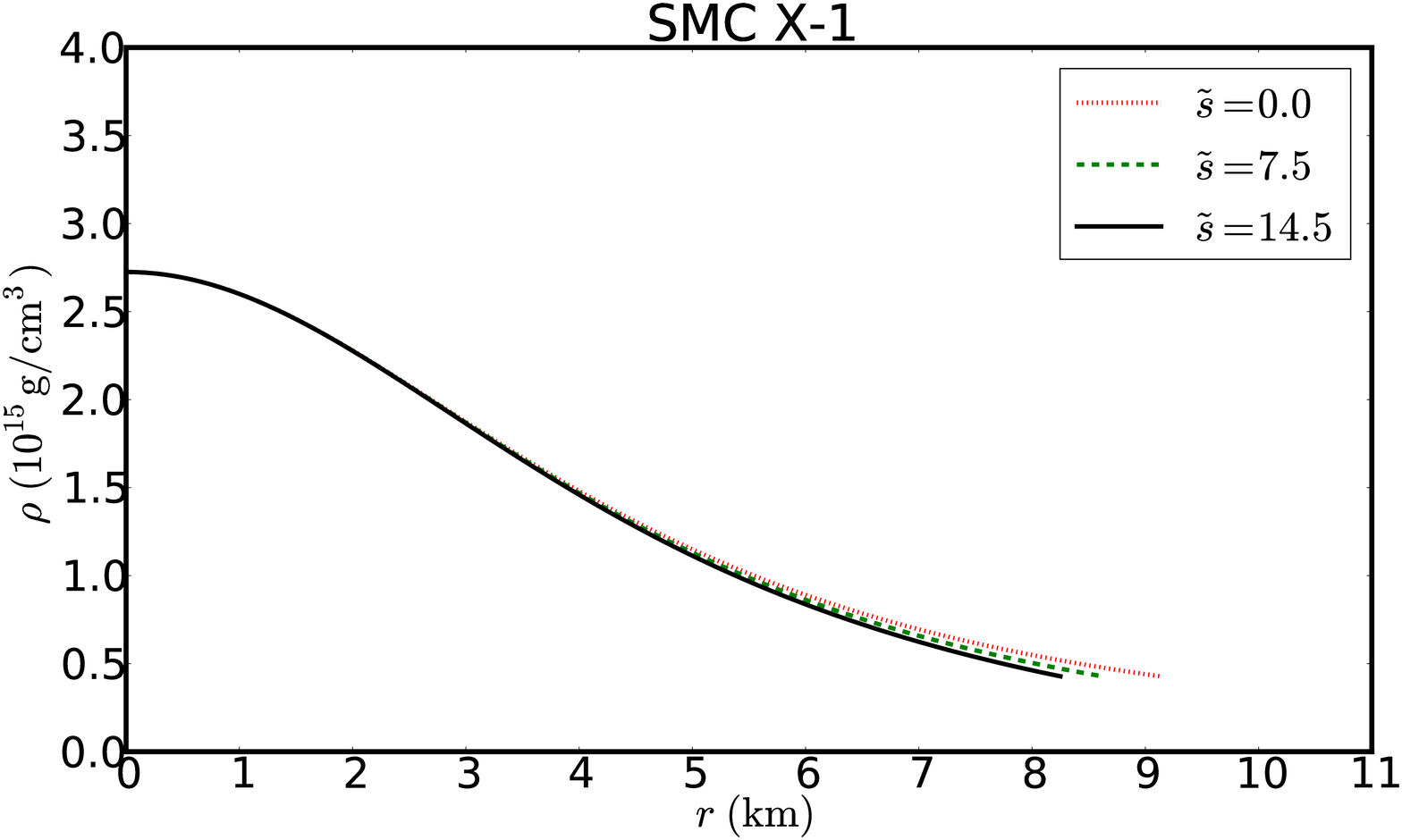} &
\includegraphics[width=0.52\textwidth]{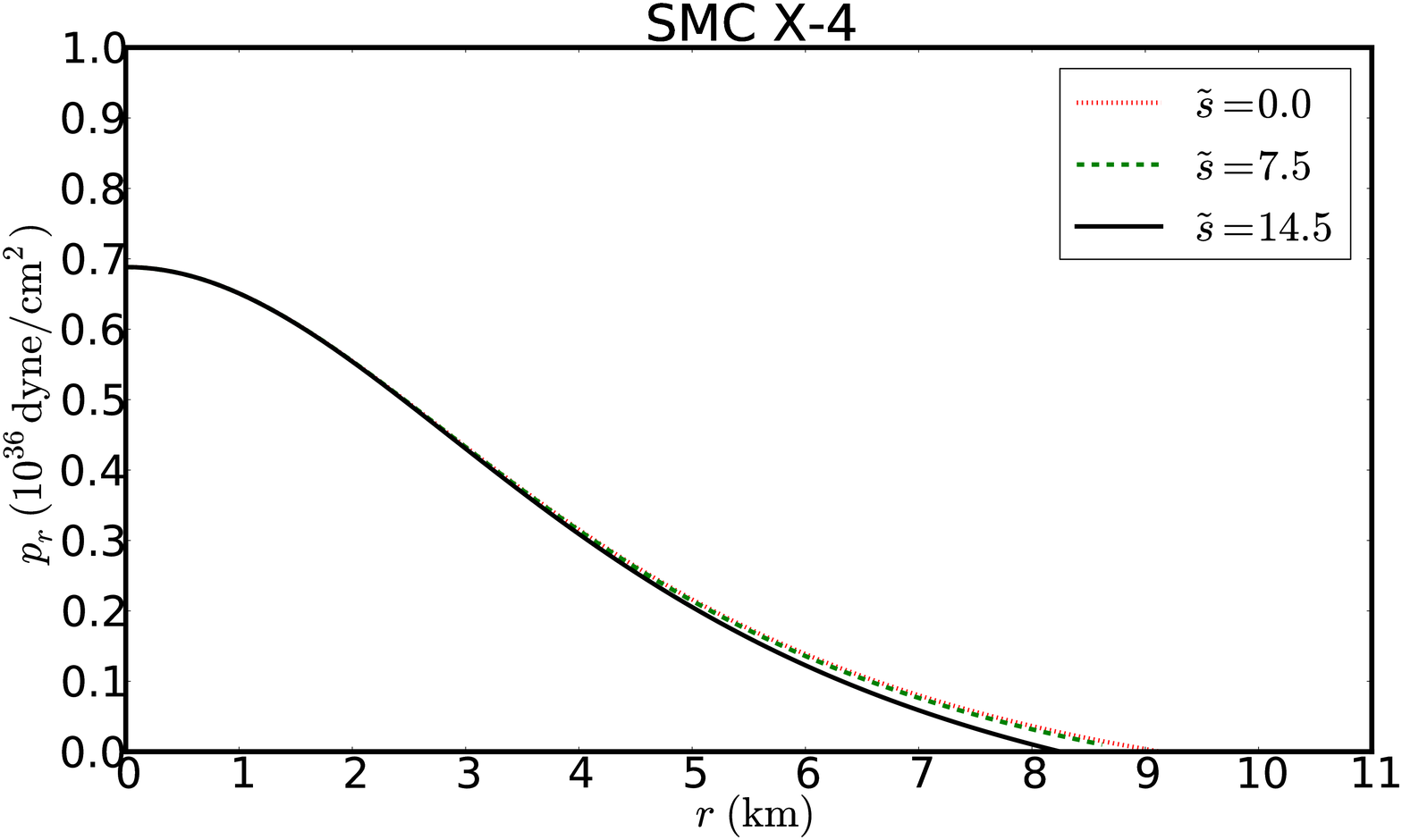}
\end{tabular}
\begin{tabular}{cc}
\includegraphics[width=0.52\textwidth]{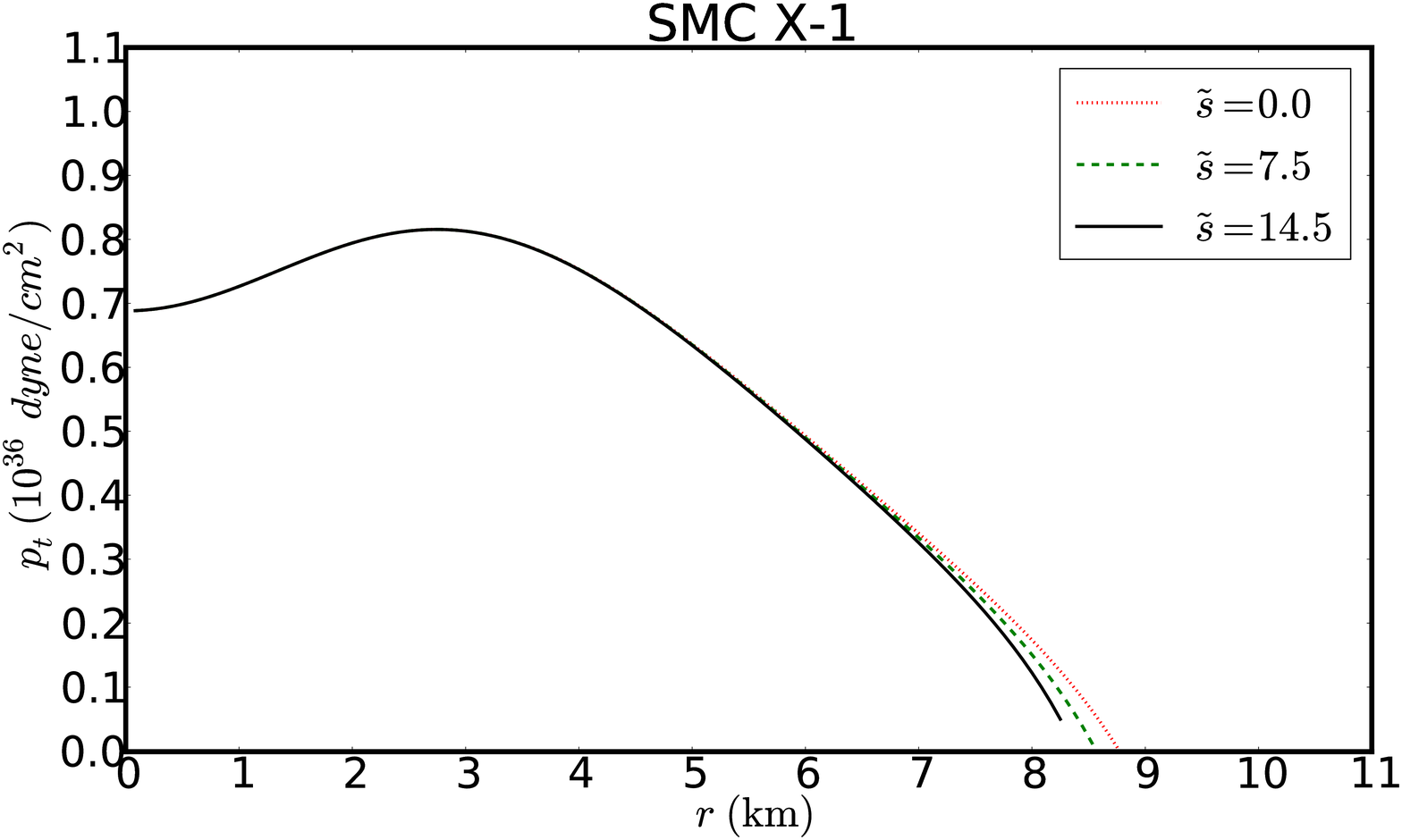} &
\includegraphics[width=0.52\textwidth]{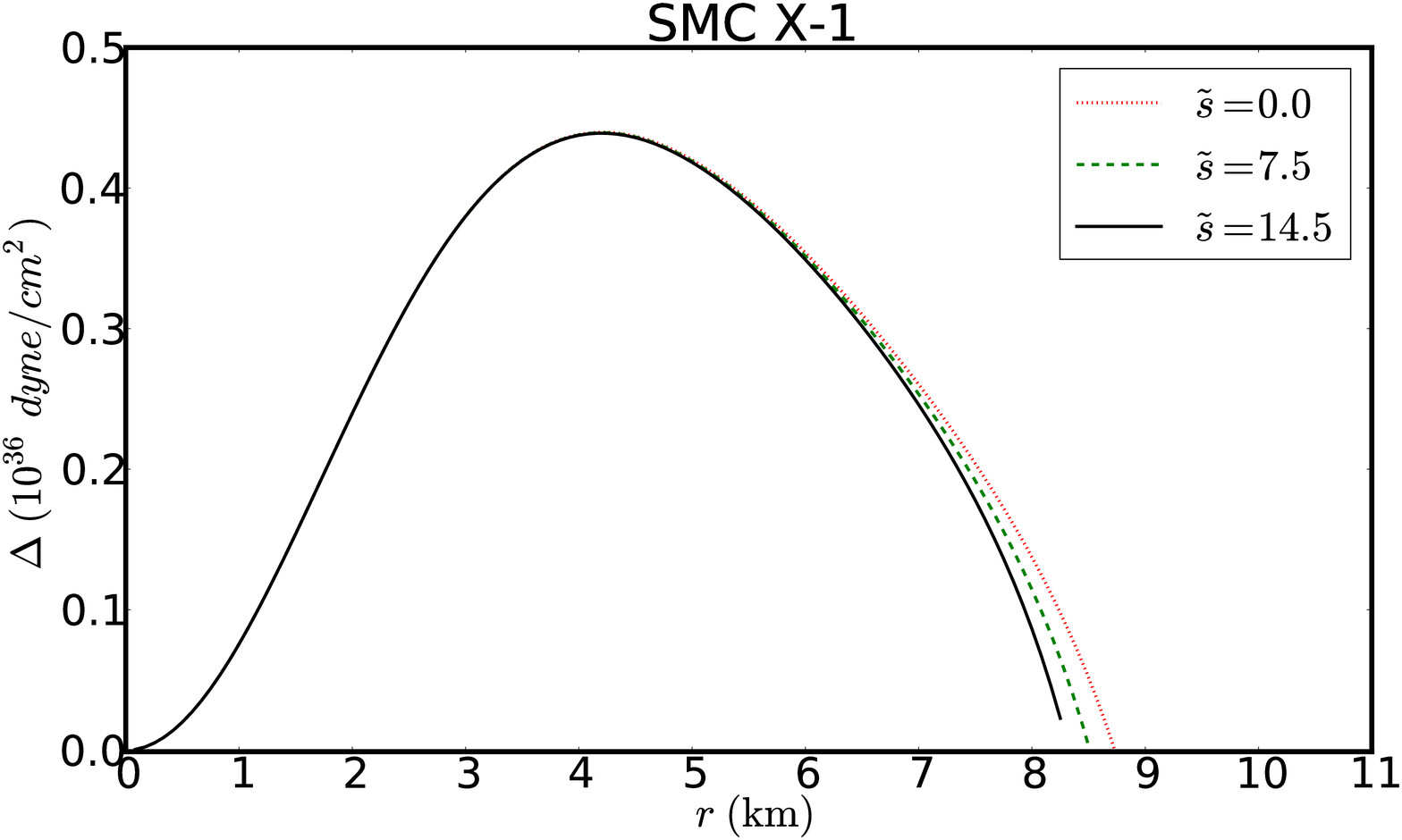}
\end{tabular}
\caption{SMC X-1, for the uncharged and charged cases.  }
\label{fig:SMC X-1}
\end{figure*} 

\begin{figure*}
\centering
\begin{tabular}{cc}
\includegraphics[width=0.52\textwidth]{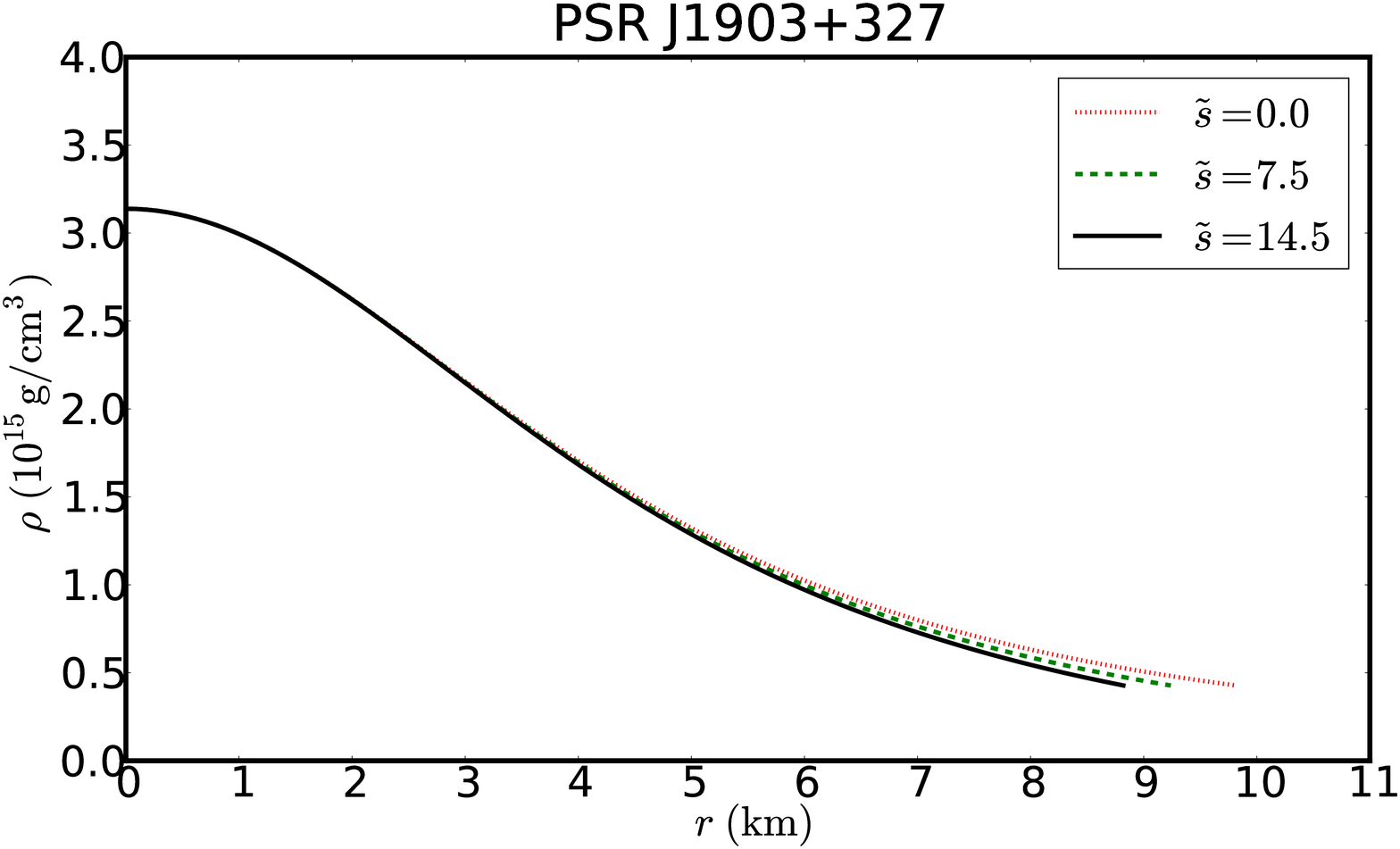} &
\includegraphics[width=0.52\textwidth]{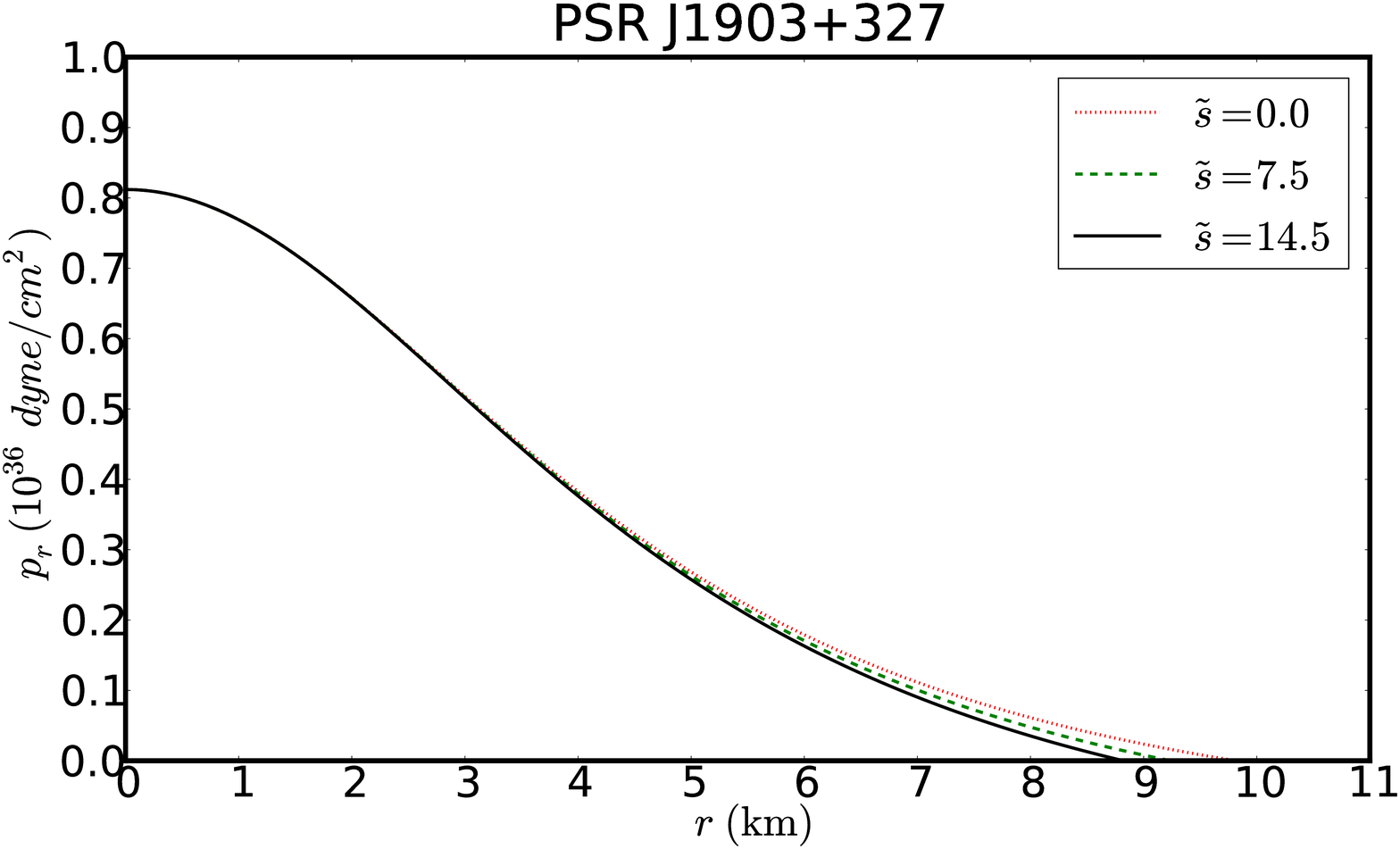}
\end{tabular}
\begin{tabular}{cc}
\includegraphics[width=0.52\textwidth]{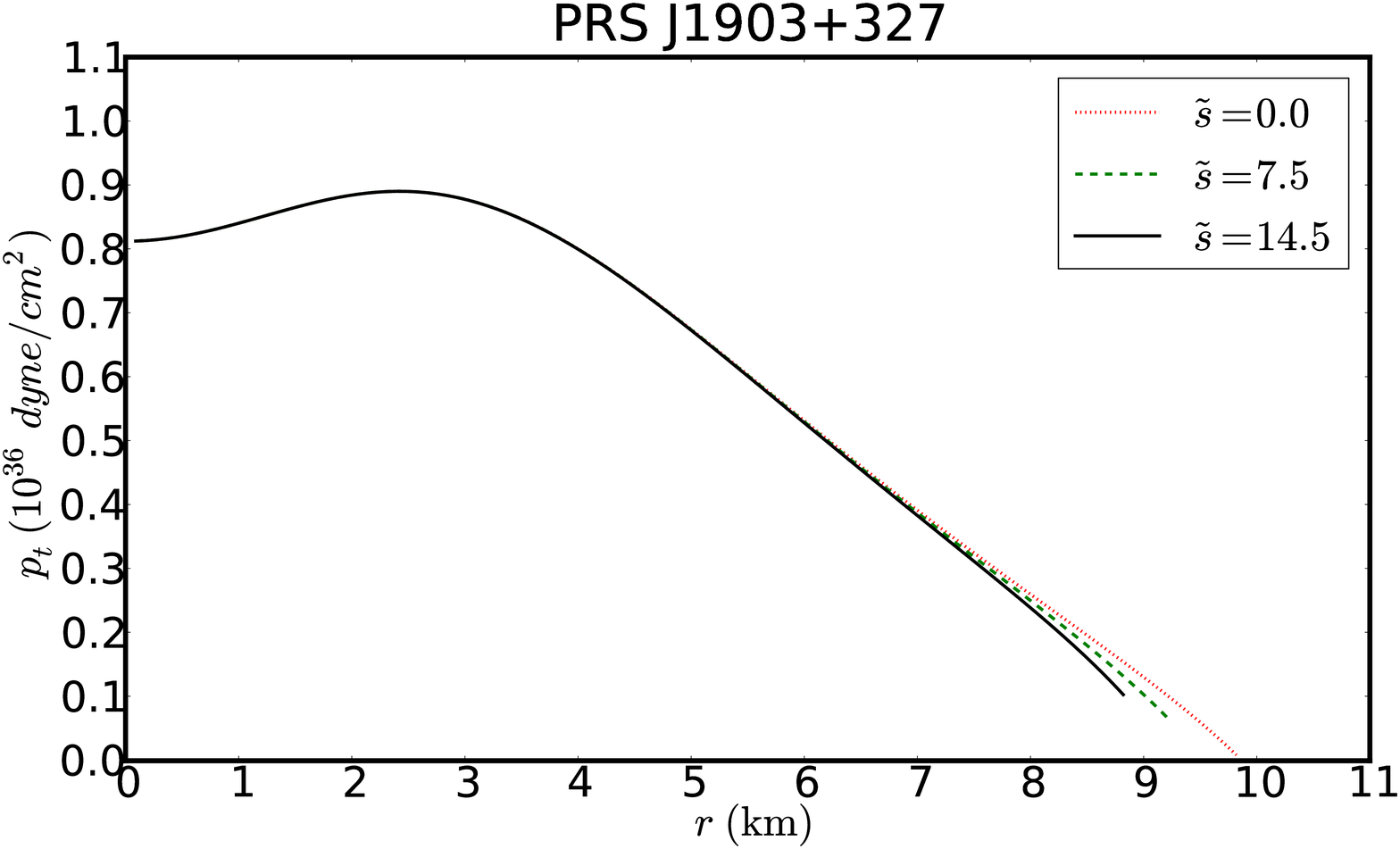} &
\includegraphics[width=0.52\textwidth]{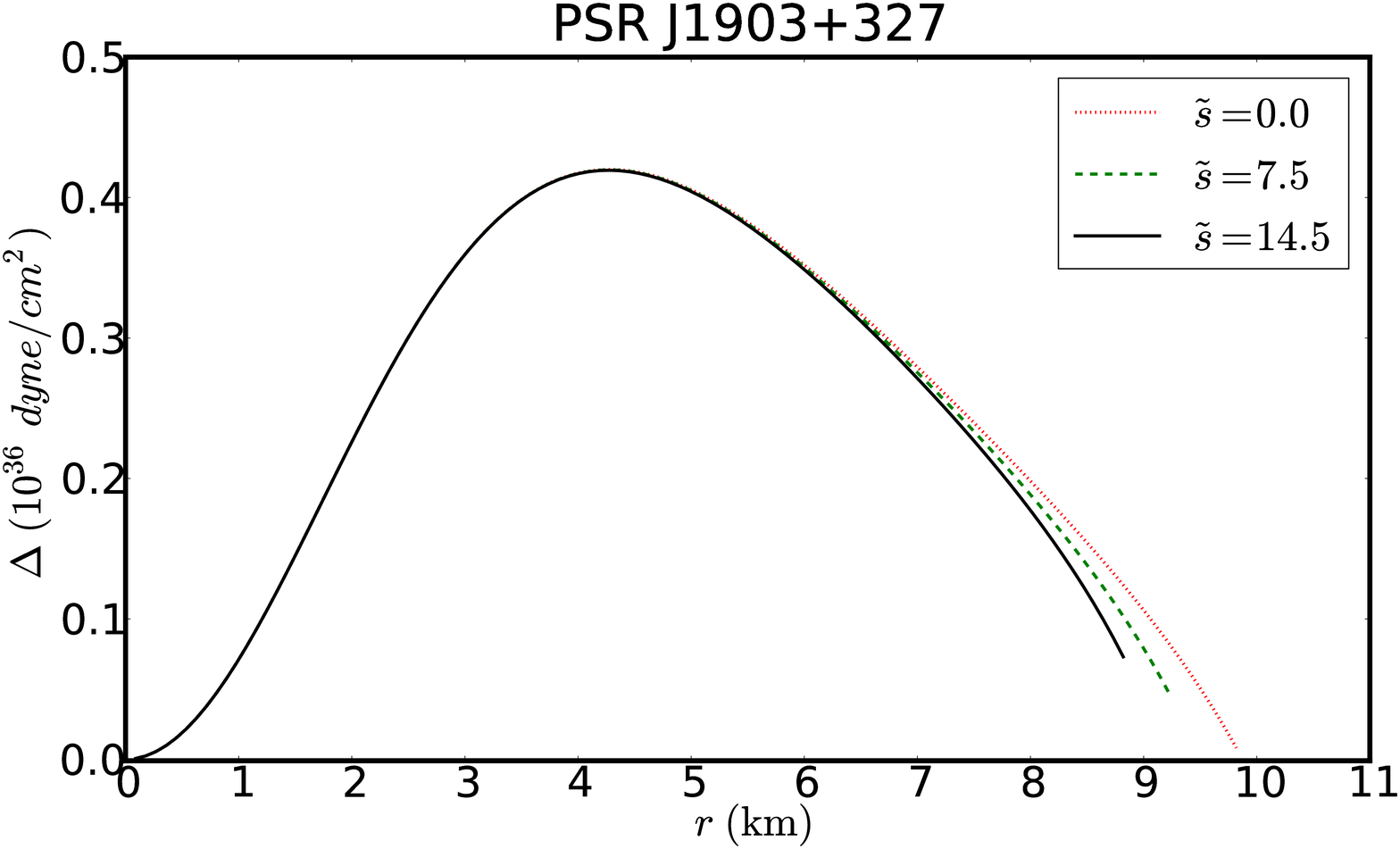}
\end{tabular}
\caption{PSRJ1903+327, for the uncharged and charged cases.}
\label{fig:PSRJ1903+327}
\end{figure*} 

\begin{figure*}
\centering
\begin{tabular}{cc}
\includegraphics[width=0.52\textwidth]{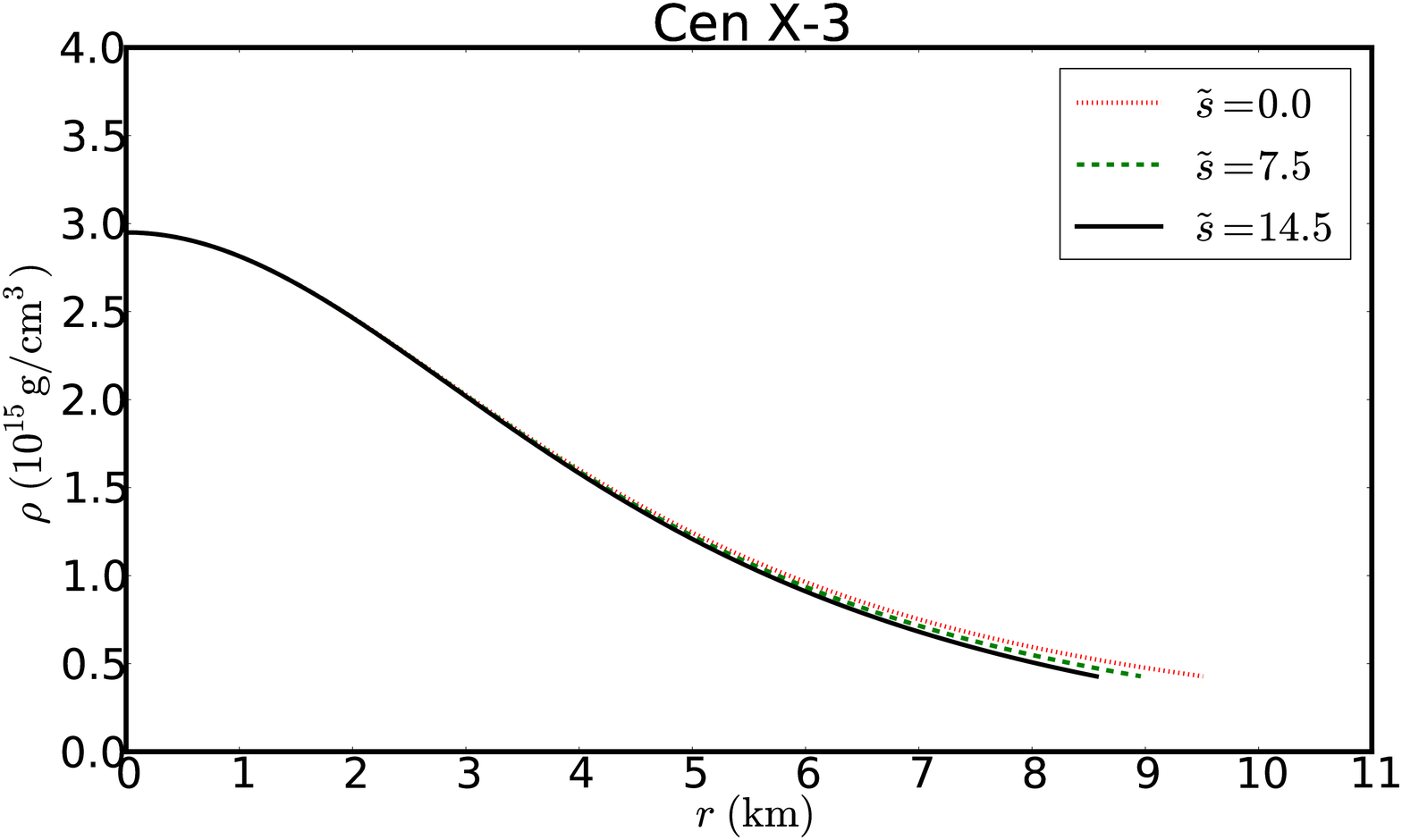} &
\includegraphics[width=0.52\textwidth]{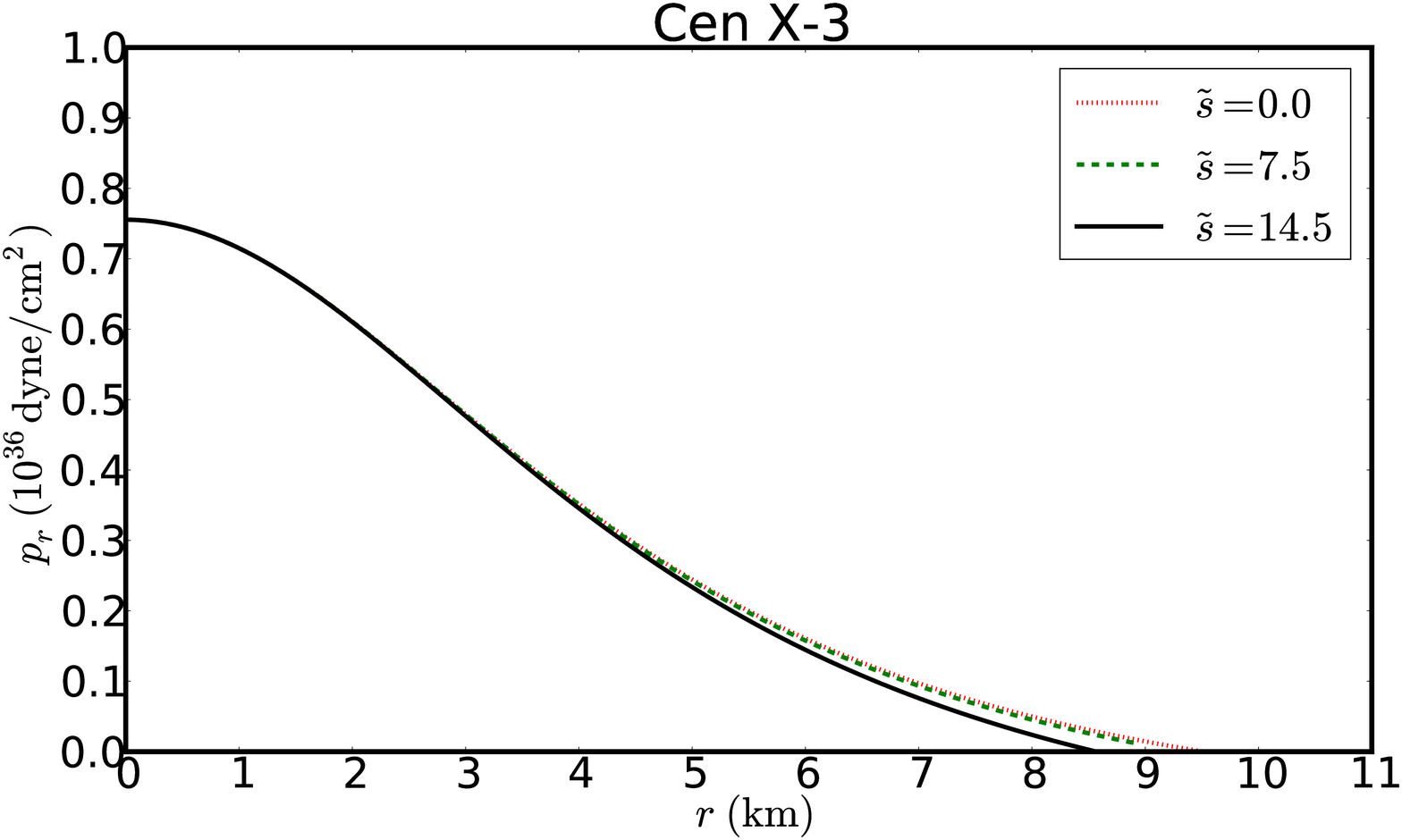}
\end{tabular}
\begin{tabular}{cc}
\includegraphics[width=0.52\textwidth]{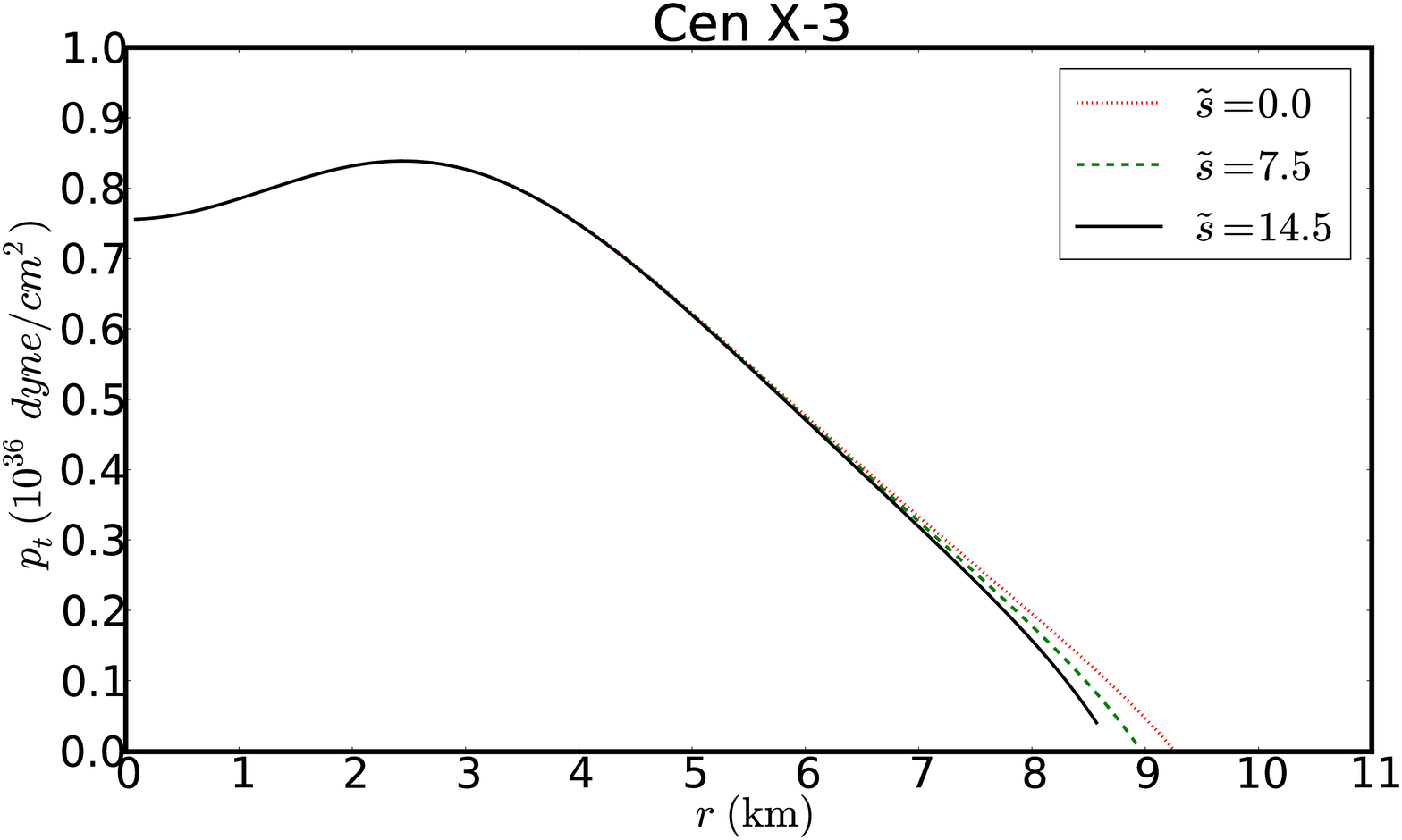} &
\includegraphics[width=0.52\textwidth]{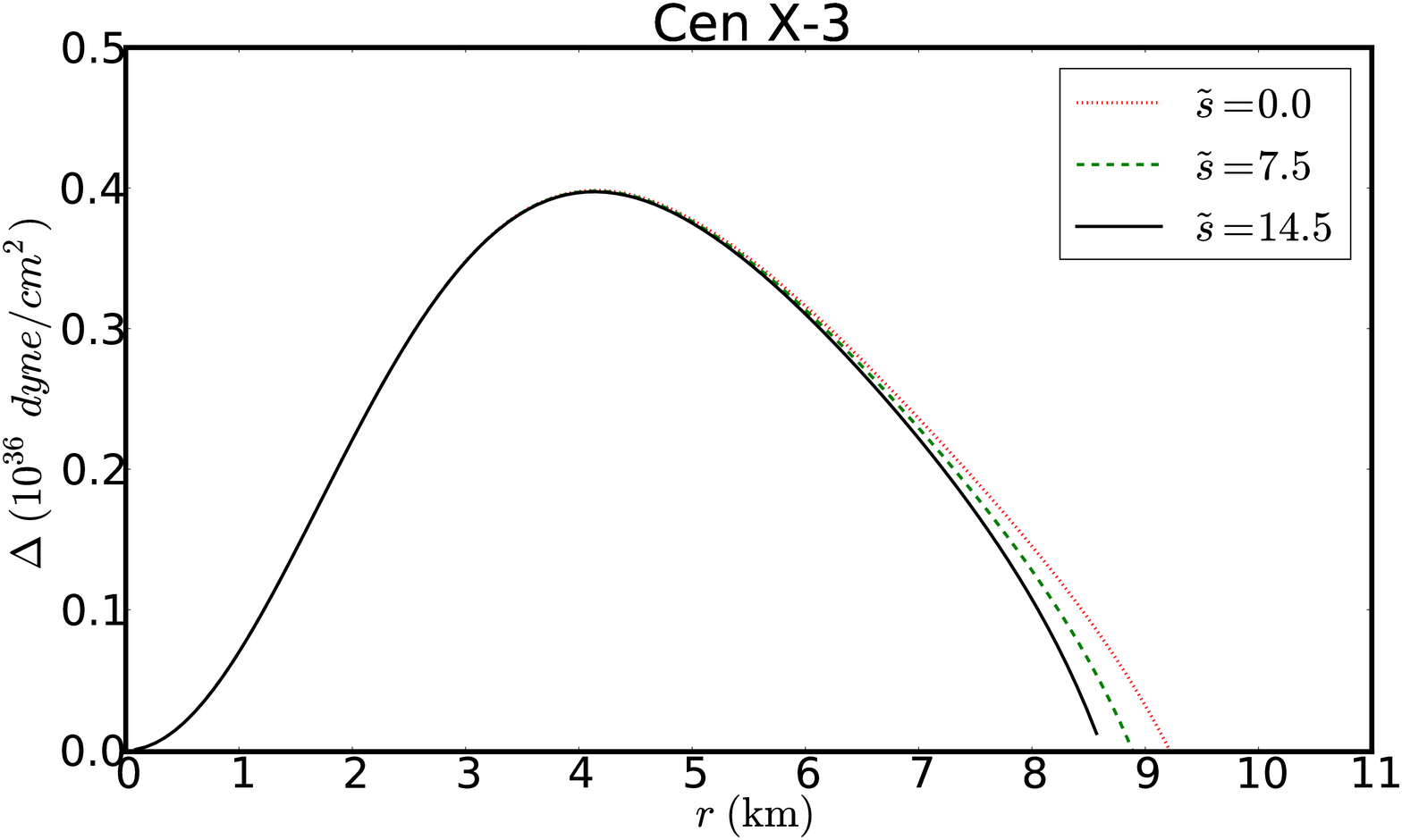}
\end{tabular}
\caption{Cen X-3, for the uncharged and charged cases. }
\label{fig:Cen X-3}
\end{figure*} 

\begin{figure*}
\centering
\begin{tabular}{cc}
\includegraphics[width=0.52\textwidth]{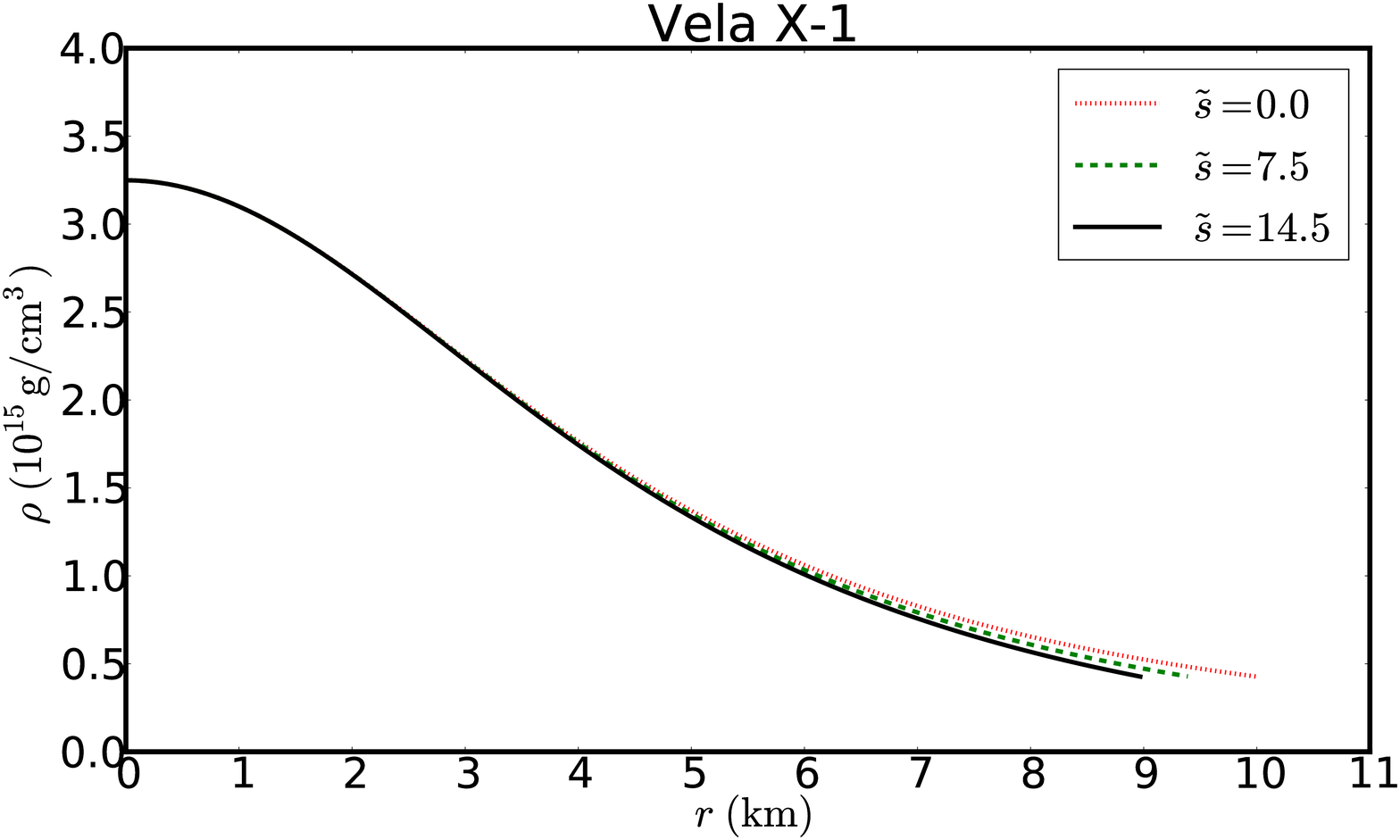} &
\includegraphics[width=0.52\textwidth]{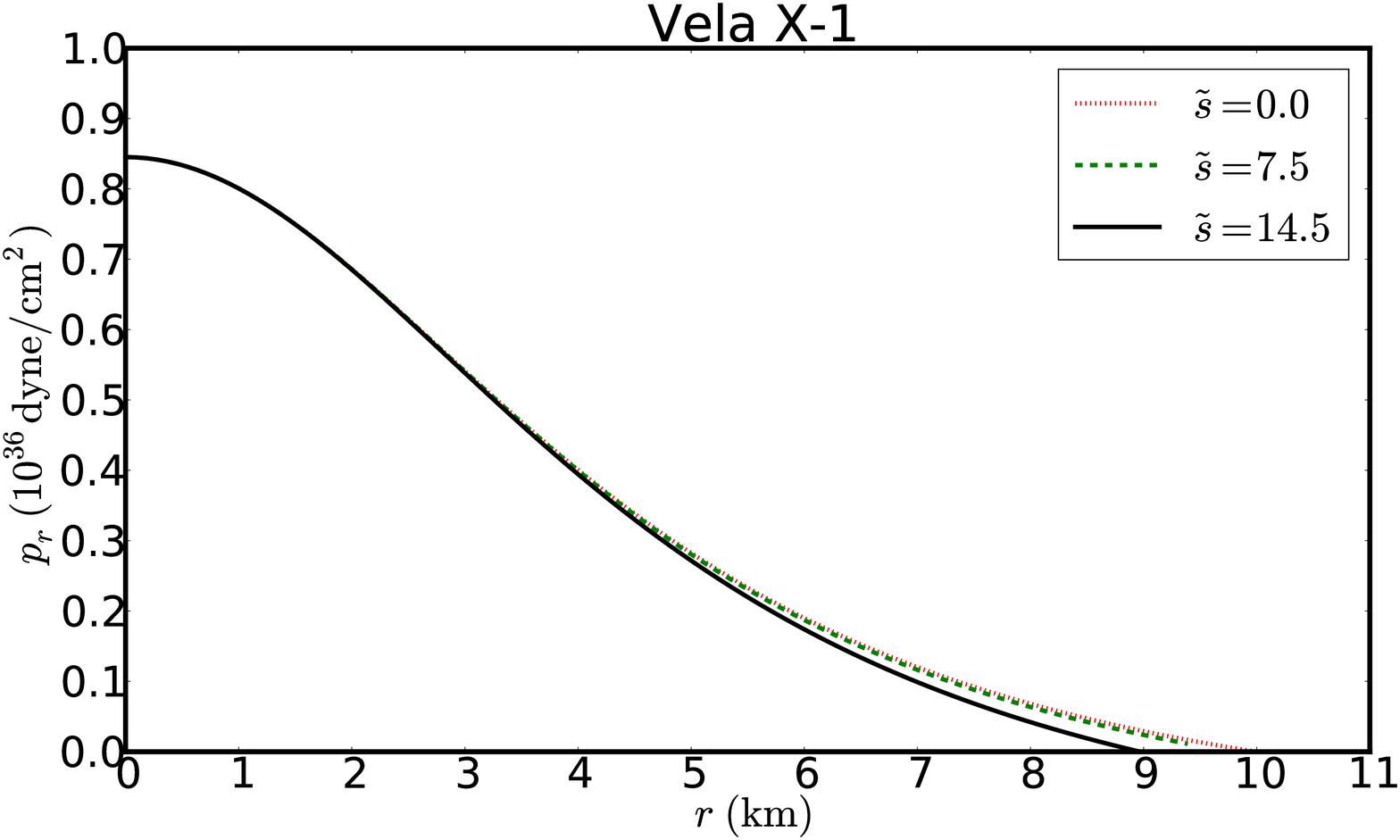}
\end{tabular}
\begin{tabular}{cc}
\includegraphics[width=0.52\textwidth]{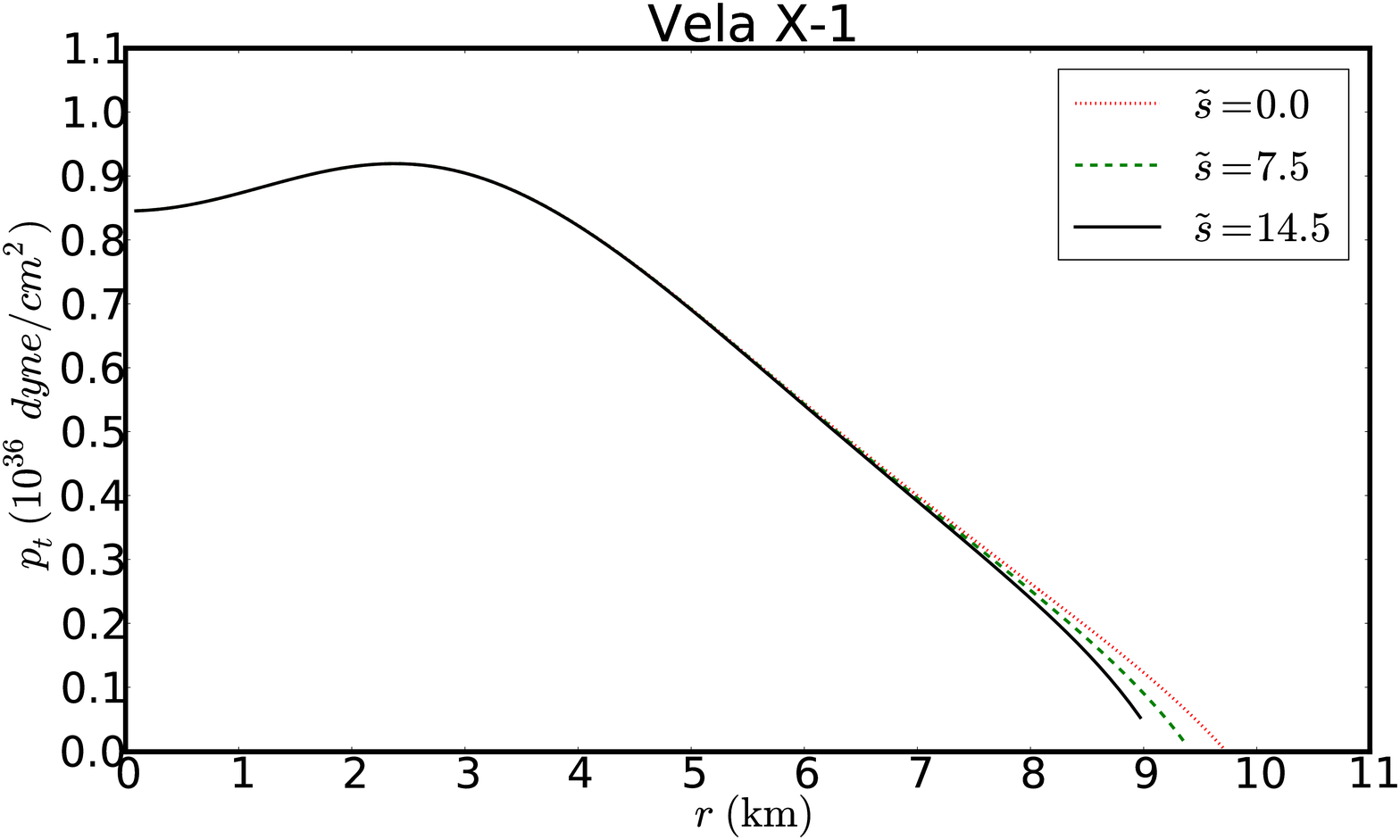} &
\includegraphics[width=0.52\textwidth]{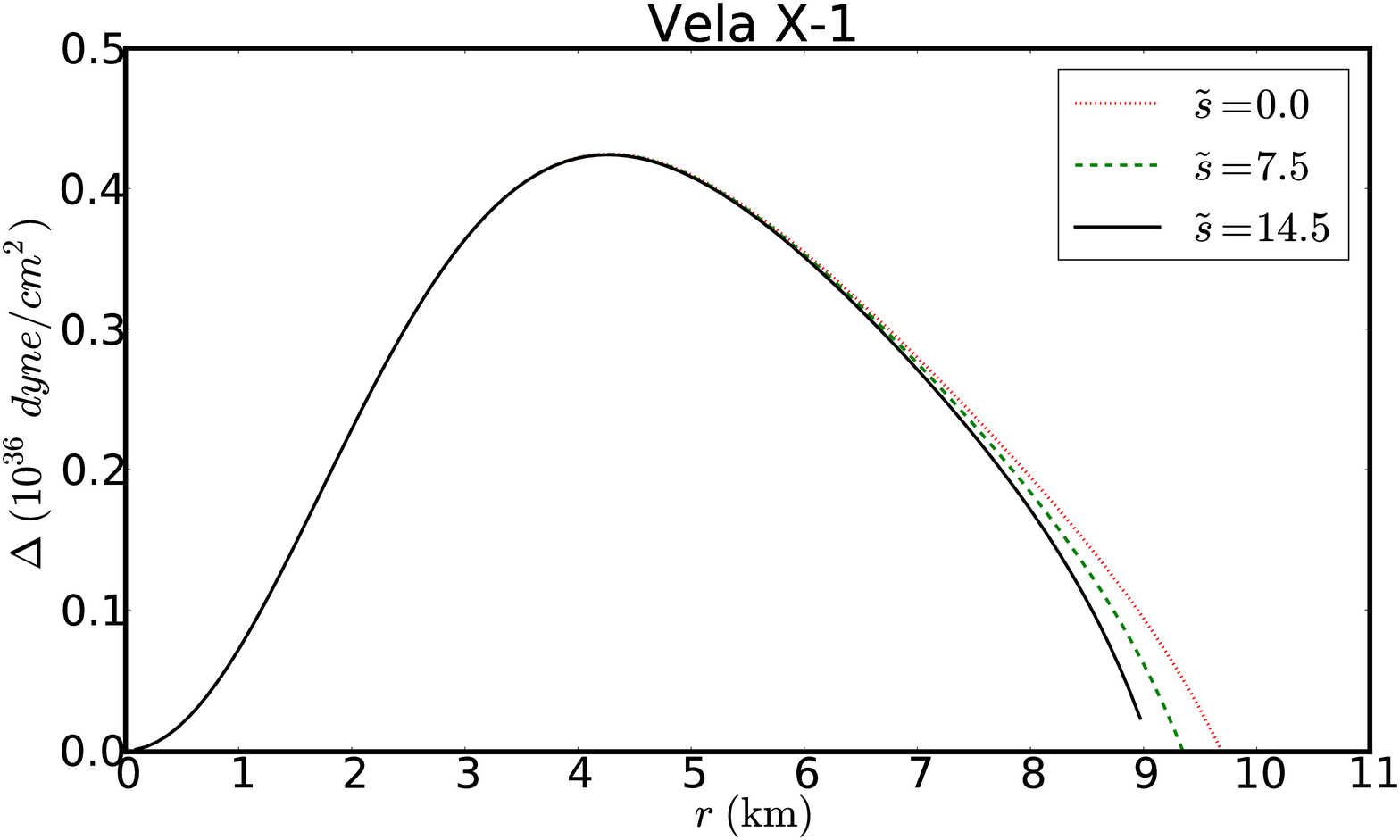}
\end{tabular}
\caption{Vela X-1, for the uncharged and charged cases.}
\label{fig:vela X-1}
\end{figure*}

\section{Uncharged stars}\label{Sec:uncharged}
In this section, we use the analytical solutions (\ref{S10a})-(\ref{S10i}) to calculate the mass and radius of five different compact stars
(PSR J1614-2230, PSR J1903+327, Vela X-1, SMC X-1, Cen X-3), and compare the output
to the recent accurate maximum mass values as mentioned in the previous section. We consider the equation of state of strange stars and
choose the central density $\rho_{c}$ in the range
of $2.2\times 10^{15}{\rm g}~{\rm cm}^{-3}\leq \rho_{c} \leq 5.5\times 10^{15}{\rm g}~{\rm cm}^{-3}$, $\tilde{a}=53.34$, $\Re=43.245~ {\rm km}$,
$\alpha=0.33$,  $\rho_{\varepsilon}=0.5\times 10^{15}{\rm g}~{\rm cm}^{-3}$. Then the model yields the above stars which are
represented in Table 1. We regain the accurate mass and corresponding radius for each star.
For PSR J1614-2230, with  $\rho_{c}=3.45\times 10^{15}{\rm g}~{\rm cm}^{-3}$, leads
to  $M=1.97M_\odot$ and $R=10.30~ {\rm km}$. For PSR J1903+327, $\rho_{c}=3.14\times 10^{15}{\rm g}~{\rm cm}^{-3}$  and we obtain $M=1.667M_\odot$
and $R= 9.82~ {\rm km}$.
By taking $\rho_{c}=3.25\times 10^{15}{\rm g}~{\rm cm}^{-3}$, we get the mass and radius of Vela X-1 $M=1.77M_\odot$ and $R= 9.99~ {\rm km}$.
The value $2.72\times 10^{15}~{\rm g}~{\rm cm}^{-3}$ leads to $M=1.29M_\odot$ and $R= 9.13~ {\rm km}$ which corresponds to SMC X-1.
Finally for Cen X-3, we
take $\rho_{c}=2.95\times 10^{15}~{\rm g}~{\rm cm}^{-3}$, and obtain the accurate mass and the radius ($M=1.49M_\odot$ and $R= 9.51~ {\rm km}$).

\begin{table}
\caption{Masses of different stars and radius for uncharged case $(\tilde{s}=0.0)$.}
\begin{tabular}{lcccccc }
\hline
STAR &$\alpha$&$\tilde{b}$ &$M(M_\odot)$ &$M(M_\odot)/R$&$R$ & $\rho_{c}$\\
& & & & &(km)& $(\times 10^{15}~\mbox{gcm}^{-3})$\\
\hline
PSR J1614-2230& 0.33& 40.11&1.97&0.191  & 10.30& 3.45  \\
PSR J1903+327& 0.33&36.48 &1.667 & 0.170 &9.82 & 3.14   \\
Vela X-1&0.33 &37.77 &1.77& 0.177  &9.99 & 3.25  \\
SMC X-1&0.33 &31.68 &1.29 &0.141 &9.13 &  2.72 \\
Cen X-3& 0.33&34.29 &1.49 &0.157  &9.51&  2.95  \\
\hline
\end{tabular}
\end{table}

\section{Charged stars}\label{Sec:charged}
The parameters used in Sect. \ref{Sec:uncharged} have generated results which are consistent with observational data. Consequently, we use these values to study charged bodies.
We take the central density in the range
of $2.2\times 10^{15}{\rm g}~{\rm cm}^{-3}\leq \rho \leq 5.5\times 10^{15}{\rm g}~{\rm cm}^{-3}$, $\tilde{a}=53.34$, $\Re=43.245~ {\rm km}$,
$\alpha=0.33$,  $\rho_{\varepsilon}=0.5\times 10^{15}{\rm g}~{\rm cm}^{-3}$, $\tilde{s}=0.0, 7.5, 14.5$ and $M_\odot=1.477$. Then the corresponding results are
given in Table 2. It is clear that the presence of electric charge leads to a considerable increase in the mass of a stellar object obeying the linear equation of state.
On the other hand the radius of different charged conﬁgurations ($\tilde{s}= 7.5, 14.5$) is smaller than the maximum radius of
the uncharged case ($\tilde{s}=0.0$). A similar situation arises in the analysis of  \cite{mak2004}.

To illustrate the behavior of physical parameters at the interior of different stars, we have plotted the energy density $\rho$, radial pressure $p_{r}$,
tangential pressure  $p_{t}$ and the measure of anisotropy $\Delta$. Figures 1, 2, 3, 4, 5 represent PSR J1614-2230, SMC X-1, PSR J1903+327,
Cen X-3 and Vela X-1 respectively.
The density profiles are positive and well behaved inside all stars. The effect of electric charge is more significant near the surface of stars;
this situation is consistent with the form of the electric field of \cite{takisa2013} in (\ref{S10g}) which vanishes at the centre $E(0)=0$.
We note that the interior profile of radial pressure  $p_{r}$, tangential pressure  $p_{r}$ and the measure of anisotropy  ${\Delta}$
profiles of PSR J1614-2230, PSR J1903+327, Vela X-1, SMC X-1 and Cen X-3 stars are completely unaffected by the electric charge layer, since
the latter is located in a thin, spherical shell close the surface. A similar statement has also been made by  \cite{negreiros2009}.
The tangential pressure  $p_{t}$ profiles for all studied stars are well behaved, increasing in the vicinity of the centre, reaches a maximum, and becomes
a decreasing function. This is reasonable since the conservation of angular momentum during the quasi-equilibrium
contraction of a massive body should lead to high values of  $p_{t}$  in central regions of the star, as pointed out by  \cite{karmakar2007}.
The anisotropy is increasing in the neighborhood of the centre, reaches a maximum value, then starts decreasing up to the boundary. The anisotropy profile is similar to
the model of  \cite{sharma2007a}.

\begin{table}
\caption{Masses of different stars and radius for charged case ($\tilde{s}\neq0$). For $\tilde{s}=0.1$, the results are similar to 
the uncharged case $\tilde{s}=0.0$. A little difference appears at  $\tilde{s}=7.5$ and the effect of charge becomes significant at  $\tilde{s}=14.5$.}
\begin{tabular}{clccccc} 
\hline
Charge parameter&STAR & & &$M$ & $R$ & $\rho_{c}$\\
&&$\alpha$ &$\tilde{b}$ &($M_\odot$) & (km)& $(\times 10^{15}~\mbox{gcm}^{-3})$\\
\hline
&PSR J1614-2230& 0.33& 40.11&1.97  & 10.30& 3.45  \\
&PSR J1903+327& 0.33&36.48 &1.667&  9.82 & 3.14   \\
$\tilde{s}=0.1$&Vela X-1&0.33 &37.77 &1.77& 9.99 & 3.25  \\
&SMC X-1&0.33 &31.68 &1.29 & 9.13 &  2.72 \\
&Cen X-3& 0.33&34.29 &1.49 & 9.51&  2.95  \\
\hline
&PSR J1614-2230& 0.33& 40.11&1.98  & 9.67& 3.45  \\
&PSR J1903+327& 0.33&36.48 &1.674&  9.24 & 3.14   \\
$\tilde{s}=7.5$&Vela X-1&0.33 &37.77 &1.78& 9.39 & 3.25  \\
&SMC X-1&0.33 &31.68 &1.30 & 8.62 &  2.72 \\
&Cen X-3& 0.33&34.29 &1.50 & 8.96&  2.95  \\
\hline
&PSR J1614-2230& 0.33& 40.11&2.13  & 9.21& 3.45  \\
&PSR J1903+327& 0.33&36.48 &1.81&  8.82 & 3.14   \\
$\tilde{s}=14.5$&Vela X-1&0.33 &37.77 &1.92& 8.96 & 3.25  \\
&SMC X-1&0.33 &31.68 &1.40 &8.25&  2.72 \\
&Cen X-3& 0.33&34.29 &1.62 & 8.57&  2.95  \\
\hline
\end{tabular}
\end{table}

\section{Discussion}\label{Sec:conc}
We have used the  \cite{takisa2013} result to model compact stars.
In our investigation, we have considered a constant slope $\alpha=1/3$ in the equation of state, and the surface density $\rho_{s}=0.5\times 10^{15}{\rm g}~{\rm cm}^{-3}$.
The surface density chosen in this work is approximately close to $4B=0.45\times 10^{15}{\rm g}~{\rm cm}^{-3}$ of  \cite{alcock1986}.
It shows that, for particular parameters values, the model can be used to describe the observed compact stars (PSR J1614-2230, PSR J1903+327, Vela X-1, SMC X-1, Cen X-3).
The recent measurement of the mass of PSR J1614-2230 provides one of the strongest observational constraints on the equation of state thus far. In our present result we have found the mass value of $M=1.97M_\odot$, $\rho_{c}=3.45\times 10^{15}{\rm g}~{\rm cm}^{-3}$ and $R=10.30~{\rm km}$ as the corresponding radius for the pulsar PSR J1614-2230.  As accurate and reliable radius measurements of this star are not yet available, our theoretical result may be useful in future investigations.
From the general relativistic structure equations and according to  \cite{buchdahl1959}, the maximum allowable compactness (mass-radius ratio) for an uncharged star is set
by $\frac{2M}{R} <\frac{8}{9}$.
The compactness values for all stars shown in Table 1, shows the acceptability of our model. Unlike others models, for example the 
models of  \cite{thirukkanesh2008} and \cite{takisa2013}, the masses for the charged case $\tilde{s}\neq0$ increases.
For the maximum charge case with $\tilde{s}=14.5$, it has been observed that our class of solutions gives us a maximum mass for PSR J1614-2230 of $M=2.13M_\odot$, with
electric field $E=4.91059\times 10^{20}~{\rm V/m}$.
This translates to an increase of 10$\%$ of mass with charge. Our results are in agreement with the work done by  \cite{negreiros2009}, who have demonstrated that the presence of electric fields of similar magnitude,
generated by charge distributions located near the surface of strange quark stars, may increase the stellar mass by up to 15$\%$; this
helps in the interpretation of massive compact stars, with masses of around $M=2.0M_\odot$.
We conclude by pointing out that such solutions may be used to construct a suitable model of a superdense object both with both uncharged  and charged matter.

\acknowledgements
PMT thanks the National Research Foundation and the University of
KwaZulu-Natal for financial support. SR acknowledges the NRF incentive funding for research support.
SDM acknowledges that this work is based upon research supported by the South African Research
Chair Initiative of the Department of Science and
Technology and the National Research Foundation.

\end{document}